\documentclass[12pt]{iopart}

\usepackage{graphicx}
\usepackage{deluxetable}
\usepackage{epsfig}
\usepackage{multirow}

\begin{document}

\title[Ancient planetary systems orbit white dwarf stars]{Ancient planetary systems are orbiting a large fraction of white dwarf stars  }

\author{B. Zuckerman$^1$, C. Melis$^1$\footnote{present address: Center for Astrophysics and Space Sciences, University of California, San Diego, CA 92093, USA.}, B. Klein$^1$, D. Koester$^2$ and M. Jura$^1$}

\address{$^1$Department of Physics and Astronomy, University of California, Los Angeles, CA 90095, USA}
\address{$^2$Institut fur Theoretische Physik und Astrophysik, University of Kiel, 24098 Kiel, Germany}
\eads{\mailto{ben@astro.ucla.edu}, \mailto{cmelis@ucsd.edu}, \mailto{kleinb@astro.ucla.edu}, \mailto{koester@astrophysik.uni-kiel.de}, \mailto{jura@astro.ucla.edu}}
\begin{abstract}
Infrared studies have revealed debris likely related to planet formation in orbit around  $\sim$30\% of youthful, intermediate mass, main sequence stars.  We present evidence, based on atmospheric pollution by various elements heavier than helium, that a comparable fraction of the white dwarf descendants of such main sequence stars are orbited by planetary systems.  These systems have survived, at least in part, through all stages of stellar evolution that precede the white dwarf.  During the time interval ($\sim$200 million years) that a typical polluted white dwarf in our sample has been cooling it has accreted from its planetary system the mass of one of the largest asteroids in our solar system (e.g., Vesta or Ceres).  Usually, this accreted mass will be only a fraction of the total mass of rocky material that orbits these white dwarfs; for plausible planetary system configurations we estimate that this total mass is likely to be at least equal to that of the Sun's asteroid belt, and perhaps much larger.  We report abundances of a suite of 8 elements detected in the little studied star G241-6 that we find to be among the most heavily polluted of all moderately bright white dwarfs.

\end{abstract}
\maketitle

\section{Introduction}

Because the photospheres of white dwarfs are expected to be composed entirely of hydrogen or helium, UV and optical spectra can be exquisitely sensitive probes of tiny quantities of all elements heavier than helium accreted from a source outside of the white dwarf (Koester \& Wilken 2006; Koester 2009).  During the 1990s research with the HIRES echelle spectrometer on the Keck telescope at Mauna Kea Observatory revealed that $\sim$25\% of cool DA (hydrogen atmosphere) white dwarf photospheres are polluted with elements heavier than helium (Zuckerman et al 2003; hereafter ZKRH03), notwithstanding the short photospheric residence time of such elements before they sink out of sight (Koester \& Wilken 2006).  At that time the origin of these heavy elements (hereafter Òhigh-ZÓ elements) was obscure.  During the past few years data from the Spitzer Space Observatory and ground-based observatories have established beyond any reasonable doubt that high-Z elements in the most heavily polluted white dwarfs come from asteroids or more massive rocky objects that first are tidally disrupted into a disk that orbits a white dwarf and subsequently are accreted onto the white dwarf (Jura 2003; Gansicke et al 2006; Kilic et al. 2006; von Hippel et al 2007; Farihi et al. 2009; Jura et al. 2009b; Farihi et al. 2010a; Melis et al 2010).  

Once it became clear that high-Z elements in heavily polluted white dwarfs originate in objects accreted from orbiting planetary systems, an important question arose:  what percentage of the polluted DA white dwarfs are accreting high-Z elements from a surrounding planetary system?  That is, when considering those white dwarfs with only moderate or minimal (but non-zero) pollution, were the observed high-Z elements initially carried by multiple rocky objects of only modest size that originate in an ancient planetary system (Jura 2008), or do these elements originate from an alternative source, for example the interstellar medium (Dufour et al. 2007)?   Our current study had two principal goals: (1) to determine the percentage of white dwarfs that are polluted by high-Z elements, and (2) to establish the source of these elements for the  ''run of the mill'' polluted white dwarf.  As a bonus, the sample of white dwarfs with measured abundances of multiple high-Z elements is now becoming sufficiently large that a diversity of elemental mixtures in the accreted materials is emerging, enabling an additional probe into their origin.  In Appendix B we report one such heavily polluted white dwarf, G 241-6.

The white dwarf sample in ZKRH03 consisted primarily of DA stars with effective temperature below 10,000 K.  Upon realizing that white dwarfs with helium-dominated (DB) atmospheres are often much richer sources of a wide range of detectable high-Z elements than are DA white dwarfs (e.g., Zuckerman et al. 2007; Klein et al. 2010), we initiated a Keck program to study the DB class. The present paper describes initial results from this ongoing study.  Previously Koester et al. (2005) and Voss et al. (2007) utilized the VLT to investigate the presence of high-Z elements and of hydrogen in atmospheres of DB stars.  The DB class is found at temperatures between 10,000 K and 30,000 K over which range it accounts for about 20$-$25\% of all field white dwarfs (Tremblay \& Bergeron 2008; Davis et al. 2009). 

\section{Sample selection}

To select stars for observation we examined the McCook \& Sion (1999) catalog of spectroscopically identified white dwarfs for stars with spectral class DB and V magnitude brighter than 16.3.  Those (bright) DB white dwarfs already known in 1999 to be polluted with high-Z elements are classified in McCook \& Sion as DBZ or DBAZ (where the Z indicates such elements and the A indicates the presence of some hydrogen in the helium-dominated atmosphere).  We did not observe these stars as part of the present program, five of which are listed at the end of Table 1, since we already knew that they would show high-Z elements, specifically the Ca II K-line, when observed with the Keck telescope.  (Klein et al 2010 observed GD40 as part of a HIRES project to investigate relative element abundances in heavily polluted white dwarfs.)

Based on our observational experience with DA white dwarfs (ZKRH03; Koester et al. 2005) and from theory (Koester 2009), we anticipated that the Ca II K-line (nearly always the strongest line from any high-Z element in an optical white dwarf spectrum over a wide range of T$_{eff}$) would have a larger equivalent width (EW) in cool, high-Z polluted, DBs than in hot DBs (see Section 4.1).  For this reason, our choice of DB stars was limited to temperature classes 4 and 5 in preference to the hotter DB3 stars listed in McCook \& Sion.  Given these constraints on V magnitude and temperature class in McCook \& Sion (1999), the sample of DB stars listed in Table 1 and plotted in Fig. 1 is very nearly complete.  That is, probably all DB4 and DB5 stars with V magnitude brighter than 16.3 in the 1999 version of the McCook \& Sion catalog and accessible with the Keck telescope are included in Table 1.  Such a sample should be unbiased in regard to discovery of DBZ and DBAZ stars from the overall DB class.  

To determine high-Z abundances and upper limits with the aid of our atmospheric models (Section 3) it was necessary to determine the temperature of each star in Table 1.  We found that the temperature labels 4 and 5 listed in McCook \& Sion (1999) were sometimes incorrect as may be seen from the high temperature tail of white dwarfs plotted in Fig. 1.  Therefore, we do not include the McCook \& Sion temperature class in the third column of Table 1.  Given problems with published temperature labels for DB stars and that the 1999 version of McCook \& Sion is incomplete, a final temperature ordering and accounting of all known, moderately bright, DB stars remains for future investigations.  

\section{Observations and analysis}

We used the HIRES echelle spectrometer (Vogt et al. 1994) on the Keck I telescope at Mauna Kea Observatory in Hawaii.  Spectra were obtained primarily during observing runs in May and November 2007, but a few stars were observed in 2006 or 2008.  The blue cross disperser was combined with a 1.15 arcsec slit and the full wavelength range between 3130 and 5940 \AA\ was covered with a resolution of $\sim$40,000.   Spectra were reduced using both the IRAF and MAKEE software packages.  Figure 2, a representative example of the quality of the spectra, presents the spectrum of WD1644+198 that contains both photospheric and interstellar Ca K-lines.   

We analyzed the various spectra with methods and input physics described by Koester  (2010), and earlier references cited therein.  The parameters for the stars were determined using various sources: values 
from the literature, which were obtained from optical and/or UV spectra, our own fits to spectra from the SPY project, fits to photometry from SDSS, or 
BVJH/BVRIJHK obtained from the SIMBAD database. As far as we know none of the model atmospheres used in all these procedures included high-Z elements.

For our own fits, however, models including traces of hydrogen were used, if 
the presence of hydrogen was known from observation of Balmer lines, or indicated by giving a better fit to the photometry.  For most of the DBAZ
in our survey the contribution of atmospheric hydrogen to the electron
abundance is larger than that of the high-Z elements because hydrogen is
significantly ionized and much more abundant than high-Z elements.  More
generally, the dominant contributors to the electron abundance in helium
atmosphere white dwarfs depend on temperature and relative hydrogen and
high-Z element abundances.  Whereas electrons donated by He will dominate
above ~15,500 K, for lower temperature stars with low H abundances and
especially large high-Z abundances, the latter should be considered in
model estimates of Teff and log g; Dufour et al (2010) remark that their
models for GD40 indicate that high-Z elements should be included in the
analysis.

Ca abundances by number with respect to He, based on the Ca II K-line, are presented in Table 1.  One exception is the previously little studied white dwarf G241-6 for which 9 Ca II lines were utilized (see Appendix B).  Although our principal objective was the Ca II K-line, two out of the three elements Fe, Si and Mg are detected in GD205 and GL837.1 (see Appendix B).  Eight elements heavier than helium are detected in G241-6 and a complete abundance analysis is presented in Appendix B.

In addition to the four white dwarfs mentioned in the preceding two paragraphs, we detected the Ca K-line in six additional stars and deduce that three of these lines are photospheric and three are from the intervening interstellar medium (ISM). We compare the heliocentric radial velocity of photospheric He lines with the Ca line.  If they disagree by more than the measurement uncertainties (which can occasionally be as large as 5-10 km/s because He lines sometimes have complex profiles), then we assume the K-line is interstellar rather than photospheric.  For G274-39 the He velocity is 60 km/s in good agreement with the K-line at 61 km/s.  With such large velocities there can be little doubt that the weak K-line is from the photosphere.  For WD1352+004, the K-line is so strong (EW = 112 m\AA) that it must be photospheric; Koester et al (2005b) report an even larger EW (150 m\AA) for this star.  The K-line in GD128 is moderately strong and very likely photospheric for the following reasons.  First, the stellar He velocity of -12 km/s agrees better with the Ca K-line velocity, also measured to be -12 km/s, than it does with the likely velocity of the nearby ISM.  According to data in Albert et al (1993), the latter velocity would likely be about -7 km/s (in the heliocentric frame).  While the discrepancy between the ISM and He velocity is not compelling, the strength of the K-line in GD 128 provides a second reason to attribute the line to the photosphere.  Specifically, at an estimated distance of $\sim$80 pc GD 128 lies well inside the local low density ''bubble''  in which the Sun resides; according to Sfeir et al (1999) this bubble extends at least as far as 150 pc toward both Galactic poles.  Considering the high Galactic latitude (67 deg) of GD 128, 45 m\AA\ would indeed be an exceptionally strong ISM line.

WD1403-010, 1940+374 and 2234+064 all display weak Ca II K-lines that we deem to be interstellar in origin.  The K-line in 1403-010 has an EW of 19 m\AA\ at a velocity of -11 km/s while the He in the photosphere is at 2 km/s.  For 1940+374 the K-line EW is 10 m\AA\ at a velocity of -13 km/s to be compared with the He velocity of 45 km/s.  Finally, 2234+064 has a K-line EW of 28 m\AA\ at a velocity of -4 km/s while photospheric He is at 21 km/s. 

\section{Results}

\subsection{Current mass accretion rates}

Detected high-Z elements might originate from a single parent body or a mixture of smaller accreted objects (Jura 2008).  Relating measured photospheric abundances to relative element abundances in the parent body or bodies requires knowledge of whether a given polluted star is in the  ''building-up'' , ''steady-state'', or  ''declining''  phase of accretion.  These terms are most clearly defined and visualized in a situation where only a single parent body is being accreted onto a white dwarf.  The building-up phase takes place after accretion has begun but before sufficient time has passed for high-Z elements to settle out of the white dwarf atmosphere (calculated settling times are given in Koester 2009).  During the steady-state phase, accretion of high-Z elements from an external reservoir (such as an orbiting ring of dust and gas) reaches an equilibrium with settling out of the photosphere (e.g., Equa. 6 in Koester 2009).  During the declining phase accretion slows and then ends and remaining high-Z elements settle out of the photosphere.  Additional discussion of these phases may be found in Koester (2009) and Jura et al (2009b).  

Fig. 1 and Tables 1 and 2 present the principal results of the present study.  The presence of high-Z pollution in Fig. 1 stars appears to be more apparent at higher T$_{eff}$ than at lower.  To produce a given EW in the Ca II K-line, a much larger accretion rate of high-Z elements is or was required onto a hot DB white dwarf than onto a cool one.  This is true independent of whether a given polluted star is in the steady-state or declining phase of accretion and it is also the situation during most of the building-up phase.  For example, with estimates of convection zone masses and diffusion times from Koester (2009) and model atmospheres described by Koester (2010), one can calculate that, to produce a given K-line EW in a steady-state situation, the calcium accretion rate onto a DB white dwarf must be about 100 times larger at 19,000 K than at 12,000 K.  Specifically, it can be seen in Tables 1 and 2 that upper limits to total mass accretion rates (based on assuming a steady-state) in the cooler white dwarfs in our sample are much smaller than the accretion rates characteristic of the hotter white dwarfs.  Should a white dwarf be in the building-up phase rather than the steady-state, then the temperature dependence of calcium accretion rates required to produce a given K-line EW is even more striking; for example, the rate would need to be $\sim$3000 times larger at 19,000 K than at 12,000 K.  

Because heavier elements tend to settle faster than lighter ones (Koester 2009), during the declining phase Fe would typically be under abundant relative to elements such as Mg and Si.  Therefore, if a suite of elemental abundances is measured in a given white dwarf and Fe is found to be relatively under abundant, then the declining phase would be plausible.  However, among five DBs with measured Fe abundances listed in Table 1 (see Appendix B) and settling times a few times 10$^5$ yr, with the possible exception of G241-6 none appear to be in the declining phase.  Likewise, DABZ GD 362 (settling time $\sim$10$^5$ yr, Jura et al 2009b) and the five DAZs listed below with short settling times are not deficient in Fe and thus none appear to be in the declining phase. Kilic et al (2008) note that the steady-state assumption is more likely to be correct for DAZ stars, where high-Z elements have relatively short convective zone lifetimes, than in DBZ stars where a circumstellar disk that is the source of photospheric pollution might dissipate while high-Z elements are still retained in the convective zone.

Both the building-up and declining phases can go on for a time span equal to only of order the settling time while, in principal, the steady-state phase can last for much longer.  Given that a typical settling time for DBZ stars in Table 1 is $\sim$1000 times shorter than the white dwarf cooling time, but $\sim$30\% of the stars with T$_{eff}$ $>$ 13,500 K have high-Z elements in their photospheres (Section 4.3.1), it is plausible to postulate that an approximate steady state is the dominant situation.  We note that this assumption is conservative in that it minimizes the mass of an accreted parent body (Jura et al 2009b).  

Mass accretion rates in Table 1 are based on the steady state assumption.  Tables 1 and 2 and Fig. 1 thus indicate an inverse relationship between accretion rate and cooling age $-$ the greater the cooling age the smaller the accretion rate, as anticipated in the Òleftover planetesimalÓ accretion model outlined in Section 5.2.  By contrast, if the source of photospheric pollution is accretion of interstellar matter, then, based on calculations of diffusion time scales and convection zone masses (Koester 2009), the percentage of DB stars with detectable high-Z elements should increase as effective temperature decreases, apparently the opposite of what our data show.  A virtue of the Table 1 sample is that it should be free of element-detection biases (see Section 2), however it could suffer from statistics of small numbers.  In Fig. 3 and Section 4.3.1 we combine our data set with that of Voss et al (2007) to improve the statistics and to draw quantitative conclusions about the percentage of DB white dwarfs with photospheric pollution.

Some words about how we estimated total current mass accretion rates are in order.  Of elements heavier than helium, carbon, nitrogen and neon appear to be too volatile to be accreted in significant abundances onto polluted white dwarfs (Jura 2006; Zuckerman et al 2007; Klein et al 2010).  We therefore neglect them and focus on Mg, Si, Fe, and O that are anticipated to carry most of the mass of rocky objects (e.g., Klein et al. 2010).  When the mass accretion rates of one or more of Mg, Si, Fe and O are known, as is the case for 6 of the Table 1 white dwarfs, then we use these directly when calculating a total current accretion rate (in g/s).  For most Table 1 stars only calcium abundances have been measured or upper limits obtained.  For such stars we turn Ca mass accretion rates into total mass accretion rates as follows.  We use the six DB stars with detected Mg, Fe and/or Si (see Appendix B) in combination with six additional white dwarfs with measured abundances to calculate relative average mass accretion rates for Ca, Mg, Si, and Fe.  These six additional white dwarfs are DABZ GD 362 (Zuckerman et al 2007; Jura et al 2009b) and DAZs WD0208+396, 1257+278, 1455+298, 1633+433 and 1858+393 (ZKRH03).  The measured mass accretion rates for Mg, Si, and Fe are typically a factor of 1.5 smaller relative to Ca than they would be if solar (cosmic) abundance ratios obtained (see Table 2 in Zuckerman et al. 2007 for relative solar abundances by mass from Lodders 2003).  Because the difference from cosmic is small and because of various other uncertainties in the analysis (for example, the best O abundance to use), we simply use the cosmic mass ratios, that is Mg/Ca = 10, Si/Ca = 11, and Fe/Ca = 19.  We are aware of five white dwarfs with measured abundances of these 4 elements and also oxygen (GD40, Klein et al. 2010; GD 378, Wolff et al. 2002, Desharnais et al. 2008; GALEX J193156.8+011745, Vennes et al. 2010; SDSS J073842.56+183509.6, Dufour et al. 2010; G241-6, Appendix B of the present paper).  Because the O abundance is better determined for GD40 than for G241-6, we rely on the relative mass accretion rates obtained by Klein et al. (2010) and assume an accretion rate for O equal to 1/2 the sum of the accretion rates for Ca, Fe, Si, and Mg.   Based on these assumed mass ratios, for stars where only Ca is detected, the total mass accretion rate given in Table 1 is 60 times the calculated Ca accretion rate or upper limit.  

\subsection{Total accreted mass} 

To derive total high-Z mass accreted by the Figure 3 white dwarf sample
we assume there is nothing special about the present time of observation.
Thus, if these stars had been observed, say, 10 Myr ago then about the
same percentage would appear as DBZs.
Figures 1 and 3 can then be interpreted in two limiting ways.  One can assume that DBZ white dwarfs have been accreting high-Z elements more or less continuously during the entire cooling age of the white dwarf, typically a few 100 million years for the DBZ stars in Table 1.  The total accreted mass is then the product of the mean accretion rate and the cooling age.  The median accretion rate for the 12 polluted white dwarfs in Table 1 is $\sim$10$^8$ g/s.  Their median cooling age is $\sim$2 x 10$^8$ yr, thus over this time interval the median total accreted mass would be $\sim$6 x 10$^{23}$ g, comparable to the mass of the largest solar-system asteroid Ceres (9.3 x  10$^{23}$ g).  Given that GD40 and G241-6 are currently accreting at a rate nearly three orders of magnitude larger than is WD1644+198, the dispersion in total accreted mass among all the polluted white dwarfs is probably quite large.  We note that if GD 40 or G241-6 has been accreting on average at about its current rate for its entire cooling age, then the total accreted mass would be comparable to the mass of Pluto (1.3 x 10$^{25}$ g).

The alternative extreme would be to assume that all white dwarfs are undergoing accretion, either from the interstellar medium or from a surrounding planetary system, but the white dwarf atmospheres are polluted to observable levels only $\sim$30\% of the time (see Sections 4.3.1 and 5.2).  Given the high upper limits to calculated mass accretion rates for the hotter DBs in Table 1, it is plausible that more sensitive observations will reveal the presence of calcium in some of these.  In any event, if all the white dwarfs in the table are accreting matter that is seen only 30\% of the time, then the median value for the total mass that has been accreted onto each white dwarf would be $\sim$2 x 10$^{23}$ g.

In Section 5.2 we consider a model of accretion of material from planetary systems orbiting white dwarfs.  When recurring accretion of rocky objects of small and moderate size is the relevant paradigm (Jura 2008), rather than accretion of a single massive object (as might be the case for GD 362, Jura et al 2009b), then the remaining mass in a typical white dwarf debris disk is larger than the total mass accreted to date.  Given the uncertain and perhaps varied physical nature of the mechanism(s) that perturb the orbits of rocky objects in toward the white dwarfs combined with a variety of plausible extrasolar debris disk architectures, it is not now possible to extrapolate from the accreted mass to the total mass of a debris disk in orbit around any specific white dwarf.  Here we mention briefly a few mechanisms that could be operational so as to motivate plausible ratios of accreted mass to total disk mass.

The gravitational field of a major planet often defines the edge of a debris disk, as per Neptune and the Kuiper Belt or the outermost  ''b'' planet that orbits the star HR 8799 (Marois et al 2008; Su et al 2009; Chiang et al. 2009).   In addition, through mean motion resonances, major planets can generate voids in debris disks, for example the Kirkwood gaps in the asteroid belt.   Debes \& Sigurdsson (2002) pointed out that the gravitational sphere of influence of a planet grows as the mass of the central star diminishes during the AGB and planetary nebula phases that precede the white dwarf stage.  This change by itself can destabilize previously stable material in a debris disk.  Among the possibilities are collisions of objects (asteroid-size or larger) that spread debris over a wide range of semimajor axes, inclinations, and eccentricities.  Much of such material might find itself in secularly unstable regions dominated by the gravitational field of a major planet.  During subsequent evolution of a white dwarf the Yarkovsky effect could drive additional debris belt material into orbits that are subject to orbital perturbations by the gravity field of a planet.  Thus, we anticipate that the accreted mass will typically be at least one and perhaps a few orders of magnitude smaller than the total disk debris mass.   If so, then the implied median debris disk masses are at least as large as that of the Sun's  asteroid belt ($\sim$4 times the mass of Ceres) and may, in some cases, rival that of the current Kuiper Belt ($\sim$0.1 Earth masses, e.g., Trujillo et al. 2001).  The orbital stability of outer planets themselves as a star becomes a white dwarf was considered by Duncan \& Lissauer (1998).

\subsection{Comparison with previous surveys of white dwarfs}

\subsubsection{The SPY sample of DB white dwarfs} 

Previous to the present Keck study of DB white dwarfs an extensive white dwarf survey was carried out with the UVES echelle spectrometer on the VLT.   Among the many papers resulting from that VLT  ''SPY''  survey the two most relevant to the Keck DB survey are by Koester et al (2005) and Voss et al (2007).  Results from the Keck and SPY surveys over the appropriate range of temperatures are displayed in Fig. 3.  Comparison of Figs. 1 and 3 reveals one noteworthy similarity and one noteworthy difference.  The similarity between the two surveys is the absence of calcium in any DB white dwarf with T $<$13,500 K.  The striking difference between the SPY and Keck samples is the much smaller percentage of white dwarfs with 13,500 K $<$ T$_{eff}$ $<$ 19,500 K and detected Ca II K-line in the former survey.  For the Keck sample the K-line detection rate in this temperature interval was 12/22 while for the VLT survey the detection rate was 8/38 (where we do not include the two DB + dM pairs in Table 1 of Voss et al 2007).

Some of the difference in K-line detection rate can be attributed to the lower sensitivity of the SPY survey.  The weakest K-line accessible to the SPY survey had an EW $\sim$50 m\AA, whereas the Keck survey was sensitive to lines typically 4 times weaker (Table 1).  Indeed, calcium was not detected in WD0125-236 by SPY, but was detected with Keck with an EW = 29 m\AA.   Ca K-lines were not detected in the SPY survey in 24 DB white dwarfs with 13,500 K $<$ T$_{eff}$ $<$ 19,500 K that were not observed in our Keck study.   Based on the Keck survey detection rate of photospheric Ca II K-lines with EW $<$45 m\AA\ (two out of 25), we anticipate that a Keck sensitivity survey of these 24 SPY white dwarfs would yield two additional DB stars.  Thus, a Keck sensitivity survey of the SPY sample would likely have yielded a Ca K-line detection rate of 10/38 white dwarfs with 13,500 K $<$ T$_{eff}$ $<$ 19,500 K.

In the temperature range 13,500 K $<$ T$_{eff}$ $<$ 19,500 K 30\% (16/54) of the Fig. 3 stars are polluted with high-Z elements.  Should the entire sample of 54 stars be observed with Keck sensitivity, then the polluted fraction would likely rise to 33\% (18/54).  This is the largest fraction of polluted white dwarfs reported in any survey of which we are aware (see Section 4.3.2).

Fig. 3 contains 14 DB white dwarfs with Teff $<$ 13,500 K none of which are polluted with high-Z elements.  If the probability for a random white dwarf to be polluted is 30\% (as per the temperature range 13,500 K to 19,500 K) then, all other things being equal, the probability that no star of 14 is polluted with high-Z elements is only 0.7\%.  However, as may be seen from Tables 1 and 2, things are not equal across the temperature range displayed in Fig.  3; in addition to a much smaller detection rate at lower temperatures, the indicated mass accretion rate is also much smaller.   How can this be?   One possibility might involve something having to do with the internal physics of white dwarfs that is presently not well understood, for example details of the mixing length treatment of convection.  Another possibility is an external mass accretion rate declining with cooling age, as would be anticipated if accretion is from a planetary system perturbed during the red giant rapid mass loss phase that preceded the white dwarf stage.  In addition, the standard classification scheme for helium-atmosphere white dwarfs likely contributes to the apparent paucity of polluted DB stars with effective temperatures $\sim$11,000 K.  At such low temperatures helium absorption lines are weak and sensitive observations are required for their detection.  A polluted white dwarf at about this temperature but with non-detected He lines would be classified as a warm DZ rather than a cool DBZ.  Such warm DZ stars do exist (e.g., Dufour et al. 2007; Fig. 1 in Jura et al 2009b).  

\subsubsection{Non-SPY surveys of white dwarfs} 

It would be desirable to extend consideration of high-Z element accretion rates to helium-atmosphere white dwarfs cooler than 10,000 K.  Investigations of substantial samples of cool He white dwarfs in classes DC, DQ and DZ have been carried out by Bergeron et al. (2001), Tremblay \& Bergeron (2008), Eisenstein et al. (2006), and Dufour et al. (2007), where the latter two studies relied on the SDSS data base.   Pierre Bergeron (private comm. 2009) has noted some potential problems inherent in using the SDSS to deduce the fraction of cool helium atmosphere white dwarfs that are polluted (i.e., the DZ class).  Dr. Bergeron opines that proper motion selected samples, as in Tremblay \& Bergeron (2008), should be unbiased and of the 87 helium atmosphere white dwarfs with temperature $<$11,000 K in Table 2 of Tremblay \& Bergeron, only 10 are certain and well-known DZ.

The occurrence of high-Z pollution can also be investigated with two extensive samples of white dwarfs that are much cooler (older) on average than the DB sample considered in Section 4.3.1.  One of these old samples contains both helium and hydrogen atmosphere white dwarfs (Sion et al. 2009) while the other sample is composed only of hydrogen atmosphere (DA) white dwarfs (ZKRH03).  The Sion et al. (2009) sample contains all 129 known white dwarfs within 20 pc of the Sun.  Because the primary atmospheric constituent (H or He) of some of these stars (e.g., some DC stars) is not now known, we consider the sample as a whole.  Table 2 in Sion et al. (2009) lists 11 DAZs, 9 DZs and one DBQZ.  However, we have found classification errors in the Sion et al. Table 2; these are listed in Appendix A, along with our Table 3 which is a revised version of Sion et al Table 2.  Our Table 3 indicates that only 17 of the 129 stars in the Sion et al. sample are polluted with high-Z elements.  The median cooling age of this entire sample is $\sim$2 Gyr (their Table 4).

A sensitive survey for the Ca II K-line (as per the current HIRES survey, ZKRH03, and Voss et al 2007) has not been carried out for most  of the Sion et al. (2009) stars.  Therefore, it is not now possible  to compute the median mass accretion rate for these 129 stars.  About 30\% of the Sion et al. sample is also included in the ZKRH03 sample that is somewhat younger ($\sim$1.1 Gyr).  If mass accretion rates decline with time, as we argue in the present paper, then the median mass accretion rate in the Sion sample is probably, if anything, smaller than the rate given below for the ZKRH03 survey. 

ZKRH03 investigated a sample of DA stars with temperatures mostly below 10,000 K and V magnitudes mostly brighter than 16.  However, for various reasons, a few of the ZKRH03 stars were quite hot.  In the statistics referred to below we only consider ZKRH03 stars cooler than 12,000 K. We also do not include G29-38 and G238-44 in the statistics.  Both of these stars were already known to be DAZ type before the ZKRH03 survey was carried out and they were observed by ZKRH because they are DAZ type (they are too hot to have been chosen by the usual survey criteria).  With these restrictions, the stars we consider are 75 single white dwarfs from Table 1 in ZKRH03.  In addition we give statistics for 81 stars that include six white dwarfs in common proper motion (cpm) pairs, where the DA star is cooler than 12,000 K, listed in Table 2 of ZKRH03.  We do not include either the double degenerate or the close DA+dM pairs listed in their Table 2.

Of the sample of 75 single white dwarfs delineated in the previous paragraph, 20 are DAZ, or 27\%.  If the six cpm pairs are also included, then DAZs comprise $\sim$25\% of the sample (20/81).  The median temperature of the entire sample is $\sim$8000 K and of the 20 DAZs, $\sim$8100 K.  Thus there is no noticeable age difference between the DAs and DAZs.  As mentioned above, the median cooling age of the entire sample is 1.1 Gyr (Bergeron et al. 1995).

In the following two paragraphs we compare mass accretion rates of DA stars from the ZKRH03 survey that have been cooling for $\sim$1.1 Gyr with the DB stars of the present survey that have been cooling for $\sim$200 Myr.  A conclusion is that no matter how one makes such a comparison, there is a dearth of heavy mass accretors at low T$_{eff}$.   

In the ZKRH03 sample, only 4 of 81 (5\%) are DAZ with mass accretion rates $>$10$^8$ g/s (0208+396, 1257+278, 1633+433, 1858+393).  This may be compared to the 19\% (10/54) of the Fig. 3 sample of DBs with T between 13,500 K and 19,500 K and mass accretion rates $>$10$^8$ g/s.  For the 20 ZKRH03 DAZ stars the median mass accretion rate is $\sim$10$^7$ g/s while the median rate for the 16 DBZ stars in Fig. 3 of the present paper is about 30 times larger.

The median mass accretion rate for the entire ZKRH03 DA sample is $<$10$^6$ g/s.  This is comparable to the median accretion rate (upper limit) for the 7 oldest (coolest) DBs in Table 2.  Comparison of the DA sample with Table 1 as a whole suggests, conservatively, that the median mass accretion rate onto white dwarfs of cooling age $\sim$200 Myr is an order of magnitude larger than for stars with cooling age of a Gyr.  Because high-Z elements are not detected in 75\% of the older sample, its actual median mass accretion rate might be much less than 10$^6$ g/s.

In summary, while it would be well worthwhile to increase the sample size of DB stars studied with the sensitivity of the present Keck survey so as to improve the statistics, all indications point to a substantial decrease in mean accretion rate of high-Z elements as white dwarf stars age.  This is consistent with a model of accretion of circumstellar rather than interstellar material as considered in Section 5.  Notwithstanding the overall decline of accretion with time, large accretion rates are maintained by an occasional cool white dwarf (such as the four DAZs listed two paragraphs above), even some with ages greater than a Gyr.

\section{Discussion}

As noted in the Introduction, the present study addresses two questions: (1) what percentage of white dwarfs are polluted by high-Z elements?, and (2) what is the source of these elements for the  run of the mill polluted white dwarf? 

As regards the first question, of all previous spectroscopic surveys of white dwarfs, the Keck/HIRES survey of cool DA stars by ZKRH03 yielded the highest fraction of polluted white dwarfs, about 1/4 (Section 4.3.2).  Based on Fig. 3 that combines the present survey with the previous VLT SPY survey, $\sim$30\% of DB white dwarfs with effective temperatures $>$14,000 K are polluted; as described in Section 4.3.1, a complete Keck sensitivity survey of such DB stars would likely yield a fraction with high-Z pollution of about 1/3.  Furthermore, as shown in Section 4.3.2, median mass accretion rates in such a DBZ sample are substantially larger than in a sample of cool DAZ stars.  Because upper limits to mass accretion rates at about 1/3 of the DB stars with Teff $>$14,000 K in Table 2 are larger than the median mass accretion rate of the DAZ stars (10$^7$ g/s) in the ZKRH03 survey, it is likely that the percentage of polluted DBs with Teff $>$14,000 K is actually larger than 1/3; but even with Keck/HIRES sensitivity not all are detectable in the Ca II K-line.

As noted in the Introduction, DBs comprise of order 20$-$25\% of all white
dwarfs in the temperature range 10,000 $-$ 30,000 K.  At present, statistics
of high-Z pollution of DA stars of comparable temperatures, along with
models that relate relative high-Z settling times in the DB and DA
classes, are sufficiently uncertain that the DB pollution percentage given
in the preceding paragraph cannot with assurance be applied to DAs with
T$_{eff}$ $>$10,000 K.  In addition, it is possible that DB and DA white dwarfs
orginate from main sequence stars with different birth environments (Davis et al 2009
and references therein), thus potentially impacting the character of their
surrounding planetary systems.  Because of such considerations, we
cannot claim with complete confidence that the percentage of planetary
systems we deduce for DB stars also applies to DA white dwarfs.  However,
the percentage of DAZ white dwarfs in the ZKRH03 sample is not much
smaller than the percentage of DBZs in the present sample and Aannestad et
al (1993), ZKRH03, and Farihi et al (2010b) all present arguments against
pollution of cool DZs and DAZs by accretion of material from the
interstellar medium.

In the following we consider the second question above: what is the source of high-Z elements in the DBZ and DAZ stars?

\subsection{Space velocities: accretion of interstellar material?}

If the DBZ stars were located on average either farther from Earth than the DB stars or if the DBZ stars were moving more slowly with respect to the local standard of rest (LSR) than the DB stars, then this could indicate that accretion of interstellar matter is responsible for at least a portion of the DBZ phenomenon.  The distance argument is as follows.  As noted in Section 3, the Sun is located in a low-density bubble of the interstellar medium (ISM) where accretion from the ISM onto white dwarfs is likely to be slow, at best.  Therefore, if accretion of interstellar material is responsible for a significant fraction of the DBZ stars, then we might expect to find these stars at a greater average distance from Earth than the DB stars.   The velocity argument is as follows: If photospheric pollution were coming from the ISM then one would expect that the DBZ stars would be moving more slowly with respect to the LSR than the DB stars (see Section 4.2 in ZKRH03).  Considering the 25 stars in the Keck survey (Tables 1 \& 4), we find no statistically significant difference between the distance to, or plane of the sky space motions of, the DBZ and DB stars.   

With radial velocities from HIRES spectra, white dwarf distances from parallaxes in the literature where available or derived photometrically, and proper motions from the literature, we have calculated Galactic UVW space motions for the 25 stars observed with HIRES (see Table 4). The total space velocities (square root of the sum of the squares of UVW) are 60$\pm$26 km/s and 68$\pm$36 km/s, respectively, for the ensembles of 18 DBs and 7 DBZs above the blank line in Table 1.  Thus there is no indication of any substantial difference in total space velocities for the DB and DBZ samples.  The median distance of the 7 DBZ stars is 72 pc.  In 3 x 10$^5$ yr $Ð$ a typical Ca settling time for these stars $Ð$ a white dwarf could go only $\sim$20 pc.  Thus, typically, these DBZ stars have not been outside of the local low-density bubble during times when they might have accreted the currently observed high-Z elements from the ISM.

Aannestad et al. (1993) and ZKRH03 presented arguments as to why nearby cool DZ and cool DAZ stars, respectively, cannot in general be polluted by accretion of material from the ISM.  (Similar to the ZKRH03 sample, only a small fraction of the DB stars -- as noted in Section 3, four of 25 -- show even weak ISM Ca
absorption lines.)  Recently, Farihi et al (2010b) came to a similar conclusion regarding the origin of high-Z elements in substantially more distant DZ white dwarfs in the SDSS.  Thus, the Keck DBZ sample is consistent with conclusions drawn by authors of these other investigations; we can find no evidence in support of accretion of high-Z elements from the ISM. 

\subsection{Planetary systems are the source of high-Z elements in most DBZ stars}

Previous research on polluted white dwarfs, primarily of the DA type, established that a few percent are surrounded by dusty disks (von Hippel et al 2007; Farihi et al 2009).  Such stars typically have the largest mass accretion rates of any of the known polluted white dwarfs.  A wide variety of arguments lead to the conclusion that the source of the dusty disk material are rocky objects, such as asteroids, that have been gravitationally perturbed into orbits that take them within the Roche radius of the white dwarf where they are shredded into tiny dust particles (Debes \& Sigurdsson 2002; Jura 2003).  Arguments that weigh against comets and/or the interstellar medium as the source of the dust can be found in papers by Reach et al. (2005), Jura et al. (2007a \& b), Jura et al. (2009a), and Farihi et al. (2009).

As we have seen, a much larger percentage of white dwarfs have photospheres polluted with high-Z elements than show evidence for circumstellar dust or gas.   As in the case of the dusty white dwarfs, a variety of arguments have been marshaled against both comets and the interstellar medium as the source of the pollution and in favor of tidal disruption of asteroids for most of the more heavily polluted stars; details may be found in ZKRH03, Jura (2006), Kilic \& Redfield (2007), Zuckerman et al (2007), and Farihi et al. (2009).  But these arguments supporting tidal disruption of asteroids as the source of photospheric high-Z elements in heavily polluted white dwarfs cannot be convincingly applied to most DAZ stars with only moderate or minimal pollution (for example, those listed in ZKRH03). 

We consider a model where the source of photospheric pollution is asteroids or other rocky bodies ''leftover planetesimals'' , Bennett et al. 2002) in a planetary system whose architecture has been disturbed by orbital expansion as a result of mass loss while the progenitor star of the white dwarf was on the AGB.  In such a scenario, one anticipates that, on average, systems settle down as they age and correspondingly less dynamical activity, such as gravitational perturbation of planetesimal orbits, is anticipated at late times.  In the model described by Debes \& Sigurdsson (2002) scattering of planetesimals by gas-giant planets would peak when a white dwarf is one to a few hundred million years old, an age characteristic of the white dwarfs in Table 1.   This, combined with a diminishing total mass of planetesimals as a white dwarf ages (cools), would imply that mass accretion rates onto a white dwarf should bear a positive correlation with T$_{eff}$.  Farihi et al (2009) present a plot of mass accretion rate vs. T$_{eff}$ for T$_{eff}$ ranging between 4000 and 20000 K (their Fig. 14).  Their plot weakly suggests, for T$_{eff}$ below about 7000 K, a possible decline in mass accretion rates.  However, small accretion rates onto the warmer white dwarfs in Fig. 14 would produce such small Ca II K-line EW as to often be undetectable.  Thus, this plot cannot be used to convincingly demonstrate or deny the existence of a correlation between mass accretion rate and white dwarf cooling age.

The main sequence progenitor of a typical planetary nebula has a mass $\sim$1.5 times that of the Sun (e.g., Osterbrook 1974), and an age $\sim$3 Gyr.  As indicated in Table 1, the cooling ages of DB white dwarfs are short with respect to 3 Gyr.   Because the DB stars are mostly younger than our solar system, it is not unreasonable to anticipate debris disks with masses at least as large as that of the Sun's asteroid and Kuiper belts at many, perhaps most, of the white dwarfs included in Table 1 (see also Section 5.4).  The larger objects carry most of the mass in these two debris regions.  Jura (2008) considered the fate of asteroids during the AGB phase of stellar evolution and concluded that most should survive into the white dwarf era provided that their initial radii were at least a few km, and their orbital semimajor axes were at least 3 AU when the progenitor of the AGB star was on the main sequence.

As demonstrated in Section 4.3.2 and by Table 2, both in the DB sample considered in the present paper and in comparison with previous surveys of DA and other white dwarfs, median mass accretion rates decrease with cooling age.  Therefore, in contrast to the lack of evidence in support of accretion from the ISM, the evidence is strong that much of the high-Z pollution is due to accretion of leftover planetesimals or perhaps, occasionally, planets.  Nonetheless, one should consider whether accretion of ISM material might account for some modest fraction of the polluted stars in our DB sample.  

We now estimate the minimum percentage of DB stars with T$_{eff}$ $>$14,000 K that are polluted by accretion of planetary as opposed to interstellar matter.  We make this estimate by considering the percentage of the more rapid accretors among the (cool, older) DAZ stars in the ZKRH03 survey.  In Section 4.3.2 we noted that 19\% of the DB stars in Fig. 3 with T$_{eff}$ between 13,500 K and 19,500 K have mass accretion rates $>$10$^8$ g/s, while only 4 of 81 stars in the much older ZKRH03 sample have such large accretion rates.  Of these 4, WD1257+278 has a very large Fe abundance (see Appendix B) and WD0208+396 has a very large Galactic space motion UVW (ZKRH03).  Thus, both of these white dwarfs appear to be especially unlikely candidates for (rapid) accretion from the ISM.  Thus of the 81 stars in the ZKRH03 sample, only about 2\% (2 stars) remain as plausible candidates for accretion from the interstellar medium at a rate $>$10$^8$ g/s  (notwithstanding that high-Z elements quickly settle out of the convection layer of such stars and that these DAZ stars are located in the local low density interstellar bubble; indeed, WD0208+396 is only 16 pc from Earth.).

Similarly, 9 of 81 (11\%) stars in the ZKRH03 sample have mass accretion rates larger than that of the least rapid accretor among the DBZ (i.e., WD1644+198 at 10$^{6.76}$ g/s) and also plausible high-Z abundances and UVW to perhaps be accreting rapidly from the ISM.  This 11\% may be compared with the fraction of DBZ, 33\%, expected in a Keck sensitivity survey of DB white dwarfs with T$_{eff}$ $>$ 14,000 K. 

Admitting that at least some low accretion rate DAZ stars are polluted by planetary system rather than ISM material implies that a super-Keck sensitivity survey of DBs would yield a DBZ fraction $>$33\%.  This, in combination with the statistics regarding high accretion rate DAZ stars described in the preceding two paragraphs, indicates that at least 1/4 of all DB stars with T$_{eff}$ $>$14,000 K are accreting from an orbiting planetary system and not from the ISM.   

\subsection{Accretion of hydrogen onto helium-dominated white dwarfs}

The relative amount of hydrogen and helium in the atmospheres of helium-dominated white dwarfs is not fully understood (Tremblay \& Bergeron 2008).  Jura et al (2009b) plot the mass of H in the convective envelope of He-dominated white dwarfs based on measurements of Voss et al (2007) and Dupuis et al (2007) and calculations of Koester (2009).  An increase in H mass over the temperature range from 20,000 K to 8,000 K is evident (Fig. 1 in Jura et al 2009b).  The He-dominated stars appear to be accreting H, but it is not now clear if this is from the ISM or from water-rich planetesimals, or both.  What is clear is that for DB stars polluted with high-Z elements the ratio of mass in such elements divided by that in hydrogen is usually many orders of magnitude larger than the cosmic ratio of high-Z elements to H.  Specifically (as noted in Section 4.2), the median total accreted high-Z mass in the Table 1 stars is a few times 10$^{23}$ g, while the typical convective layer mass in H is $\sim$10$^{22}$ g.

Many DBs in the Voss et al (2007) sample contain hydrogen (are DBA stars), but with no evidence of any high-Z pollution.  It is possible that some DBA stars once accreted high-Z elements along with H but the former have all settled below the convective zone.  If the H in the DBA stars in the Voss sample was initially carried by water-rich parent bodies, then an implication is that at least half of all white dwarfs are orbited by minor planets (Jura \& Xu 2010)  

\subsection{Comparison with main sequence planetary systems}

Precision radial velocities, transit, and microlensing studies have revealed various aspects of extrasolar planetary systems with semimajor axes of a few AU or less.  For example, from a study of post-main sequence giant stars, Bowler et al. (2010) report that 1/4 or more of intermediate mass, main sequence, stars have massive planetary companions within 3 AU.   Data on regions beyond a few AU comes primarily from statistical studies of debris disks orbiting main sequence stars and direct imaging of a few warm, young planets.  A relevant example of these latter systems is late-A type star HR 8799 with a similar architecture to the outer regions of the solar system.  A dusty Òasteroid beltÓ orbits interior to three massive planets that orbit interior to an outer dusty ÒKuiper BeltÓ (Marois et al 2008; Su et al 2009).  If the planets are retained for the full main sequence lifetime of HR 8799, then this star would represent a fine example of the multiple giant planet system Debes \& Sigurdsson (2002) envision could be involved in destabilization of a debris belt following mass loss on the AGB.  ÒExtreme adaptive opticsÓ instruments anticipated in the coming few years (e.g., Macintosh et al. 2008) should reveal additional examples of stars with gas giant planets and dusty debris disks.

The results presented in the present paper can be compared with those from surveys for debris disks around main sequence stars.  Almost all known main sequence debris disks have semimajor axes greater than a few AU, so that they sample a similar realm of planetary space as do the rocky objects that are the primary source of the calcium in the polluted white dwarfs in Table 1.  Both IRAS (Rhee et al 2007) and Spitzer (Su et al 2006; Trilling et al 2008) surveys of intermediate mass stars (primarily A- and F-type) indicate that the fraction of such stars with detectable debris disks declines with stellar age.  For the youngest ages, 10-30 Myr old, this fraction is often as large as 1/3 (Su et al. 2006; Rebull et al. 2008; Currie et al. 2008; Gorlova et al. 2007), but diminishes to perhaps 10-15\% for middle age stars (Su et al. 2006; Trilling et al. 2008).   Our observations of pollution in a substantially greater percentage of warm white dwarfs than 15\% indicate that debris systems are still present, albeit undetectable at the current state of the art, while an intermediate mass star is of middle age and still on the main sequence.   

Infrared measurements of main sequence debris disks probe the smallest particles in a distribution where most of the mass is contained in large objects; photospheric pollution of white dwarfs preferentially samples the large objects.  Detailed analysis of the relative abundances of elements in the polluting material (Zuckerman et al 2007; Jura et al 2009b; Klein et al 2010) provides a unique tool for measuring the bulk composition of rocky extrasolar objects.

\section{Conclusions}

In the early 1990s Dupuis and collaborators published a heroic series of three papers on a ''Study of Metal Abundance Patterns in Cool White Dwarfs'' (Dupuis et al. 1992, 1993a, 1993b) thus setting the stage for virtually all subsequent work on white dwarfs with photospheres polluted with elements heavier than helium (high-Z elements).  Contrasting the conclusions of the last of their papers (Dupuis et al 1993b) with the conclusions of the present paper illustrates the great and surprising advances that have taken place in this field.   Dupuis and colleagues (and others) at that time were focused on accretion of material from the interstellar medium (ISM).  While Dupuis et al were reasonably satisfied that the ISM was indeed the source of photospheric pollution in most cases, in a contemporaneous study Aannestad et al (1993) came to the opposite conclusion that accretion from the ISM was not important in most cases.

According to Dupuis et al (1993b), in their picture of accretion of material from the ISM one anticipates that ''As a general rule, the \{high-Z element\} detection probability is dominated by visibility effects (and not by abundance effects).  It increases monotonically É with decreasing temperature, \{increasing age\}, because the transparency of the atmosphere increases.''    In contrast, the present study reveals that the detection probability is actually dominated by abundance effects rather than visibility effects because the abundance of high-Z elements is substantially larger in the atmospheres of younger (hotter) white dwarfs.

The principal conclusions of the present study are:

1) The photospheres of $\sim$1/3 of DB white dwarfs with 13,500 K $<$ T$_{eff}$ $<$ 19,500 K are polluted with high-Z elements; such stars are classified as DBZs.  Because the present survey suffers a bit from statistics of small numbers the actual percentage of polluted stars could be somewhat larger or smaller.  However, because the quantity of mass accreted onto a hot DB star must be very large to yield a detectable Ca II K-line, it is plausible that some hot DBs will eventually be shown to contain calcium.  Therefore, 1/3 is more likely to represent an underestimate rather than overestimate of the percentage of DBs that are actually DBZs.

2) A comparison of the distribution of high-Z element pollution with cooling age of DB and DA white dwarfs indicates that at least 1/4 of DB class white dwarfs with 13,500 K $<$ T$_{eff}$ $<$ 19,500 K are undergoing accretion from orbiting planetary systems.  And, as has been demonstrated in the literature, the accretion is via tidal destruction of rocky objects of substantial size and not of the dribbling in of dust particles from debris disks with semimajor axes of many AU.  Integrated over their cooling times, the median total mass of elements heavier than helium that has been accreted onto the DBZ stars in our sample is comparable to that of the largest solar system asteroids.

3) Because the accreted mass in most cases will only be a fraction of the total mass in rocky objects orbiting the DBZ stars, this total mass should typically be at least comparable to the mass of the Sun's asteroid belt and possibly much larger.  One white dwarf that may well have a relatively massive complement of orbiting rocky material is G241-6, a little studied star that our survey has revealed to be heavily polluted with high-Z elements.

4) The (small) current sample of polluted white dwarfs with measured abundances of multiple heavy elements already hints at the wide diversity of elemental mixtures that will be revealed by deep, extensive, future surveys.  For example, a large iron abundance in some polluted white dwarfs suggests that at least some parent bodies were differentiated and thus quite massive.  White dwarf photospheres reveal the bulk elemental composition of large rocky extrasolar objects, knowledge that, typically, is not accessible for rocky objects in our solar system.

We thank Pierre Bergeron and the referee for various helpful comments.  This research is supported by grants to UCLA from NASA and the NSF.  Data presented herein were obtained at the W.M. Keck Observatory, which is operated as a scientific partnership among the California Institute of Technology, the University of California, and the National Aeronautics and Space Administration.   The Observatory was made possible by the generous financial support of the W.M. Keck Foundation.  We thank the Keck Observatory support staff for their assistance.  We recognize and acknowledge the very significant cultural role and reverence that the summit of Mauna Kea has always had within the indigenous Hawaiian community.  We are most fortunate to have the opportunity to conduct observations from this mountain.

\appendix

\section*{Appendix A: White dwarfs within 20 parsecs of the sun}

Sion et al (2009) present a list of 129 white dwarfs within 20 pc of the Sun.  While preparing the present paper we discovered various classification errors in Table 1 of Sion et al. for stars that (supposedly) display photospheric high-Z elements.  In this appendix we list the specific errors and present in our Table 3 a corrected revision of the Sion et al (2009) summary Table 2.

WD0108+277 classified as a DAZ in Table 1 of Sion et al (2009), is actually a DA (Farihi et al 2009).
WD0738-172 classified as a DAZ by Sion et al, is actually a DZA (Farihi et al 2009; Bergeron et al 2001).
WD1334+039 classified as a DZ by Sion et al, is actually a DC (Farihi et al 2009; ZKRH03; Bergeron et al 2001).
WD1917-077 classified in Table 2 of Sion et al as a DBQZ is actually a DBQA (Voss et al 2007).
WD2336-079 is classified as a DAZ by Sion et al, but there is no published evidence for the presence of the Ca II K-line (E. Sion, personal comm. 2009).  Hence, we classify the star as DA. 

\section*{Appendix B: Elemental abundance ratios in G241-6 and other polluted white dwarfs}

Table 5 presents element abundances and steady-state mass accretion rates in G241-6 for an atmospheric model with T$_{eff}$ = 15,300 K and log g set to 8.0. This temperature is based on (1) a fit to the He lines in the HIRES spectra and comparison with He lines in the spectrum of GD 40 (B. Klein et al, in preparation), and (2) assuring that the derived Mg abundance is the same independent of whether observed Mg I or Mg II lines are used (see Notes to Table 5).   We estimate an uncertainty in T$_{eff}$ of $\sim$400 K for G241-6; this value depends on the assumption that log g for GD 40 and G241-6 are similar.  As may be seen from Table 1, G241-6 is one of the most polluted known DBZ stars, only GD40 has a comparably large mass accretion rate; DABZ GD362 is the white dwarf with the largest currently known rate of mass accretion, $\sim$10$^{10}$ g/s (Jura et al 2009b).

As noted in Section 4.1, Klein et al (2010) calculated a steady-state accretion rate for oxygen that is half the sum of the accretion rates for Ca, Fe, Si and Mg.  For G241-6, Table 5 indicates an accretion rate for O that is 0.84 times the sum of the rates for these 4 other elements.  If one assumes that O is accreted onto G241-6 only in the form of oxides of elements heavier than hydrogen (see Klein et al 2010 for a detailed discussion) then it is not possible to account for a factor as large as 0.84.  Also, our derived upper limit to the H abundance in G241-6 (Table 5) indicates that very little, if any, O has been accreted onto the star in the form of water.  Because the oxygen lines are weak (Fig. 4), it remains possible that the ensemble of model uncertainties and measurement errors for all 5 elements might (just barely) allow a self-consistent, steady-state, model of accretion of metal oxides.

Detection of a dusty disk at G241-6 could be used to confirm that accretion is in the steady-state phase.  (Near-infrared JHK measurements with the Gemini camera on the
Shane 3-m telescope at Lick Observatory do not indicate any appreciable
excess emission in the K-band (C. Melis 2009, private communication)).  We note that a 72 pc distance to G241-6, derived photometrically, has been preferred in Table 4 to the published parallax distance of 67$\pm$20 pc.
Additional discussion of the pollution of G241-6 and comparison with other highly polluted white dwarfs will be forthcoming (B. Klein et al., in preparation).

Of the 25 stars listed above the blank line in Table 1 in our HIRES survey, GD205 and GL837.1 have detectable lines from elements other than Ca.  We see Si II 3856 and 3863 \AA\ in GD205, Fe II 3213 and 3228 \AA\ in GL837.1, and Mg II 4481 \AA\ in both stars.  The total mass accretion rates listed in Table 1 include the derived abundances and accretion rates based on the measured EW for these lines (rather than employing the general prescription outlined in Section 4.1 when only the Ca II K-line is detected in a star).   For the 5 stars below the blank line in Table 1 and not measured in our survey we use references listed in the Notes to the Table for abundances, when available, of elements other than Ca.  For GD408 and G200-039, Si and Fe abundances have been measured, and for GD40, the abundances of all relevant elements.

One noteworthy outcome to emerge from consideration of abundance ratios is [Fe/Mg] in GL837.1.   While the cosmic Fe to Mg abundance ratio by mass is about a factor of two, the measured ratio of steady-state mass accretion rates in GL837.1 is $\sim$10.   Measurement errors for EW are at the 20\% level in this star and the overall uncertainty in the relative abundance of these two elements is likely to be a factor of two or less.  Thus, the data suggest that iron-rich material has been accreted onto this star, perhaps from parent bodies similar to those giving rise to iron meteorites or to the core of Mercury.  WD1257+278 appears to be a DAZ white dwarf with a comparably large (over)abundance of Fe relative to Mg (ZKRH03).

\noappendix

\section*{References}
\begin{harvard}

\item[Aannestad, P., Kenyon, S., Hammond, G. \& Sion, E. 1993, AJ 105, 1033]
\item[Albert, C. et al 1993, ApJS 88, 81]
\item[Bennett, J. et al. 2002, The Cosmic Perspective, Second Edition (Addison Wesley, San Francisco)]
\item[Bergeron, P., Leggett S.K. \& Ruiz, M. T. 2001, ApJS 133, 413]
\item[Bergeron, P. Wesemael, F. \& Beauchamp, A. 1995, PASP 107, 1047]
\item[Bowler, B. et al. 2010, ApJ 709, 396]
\item[Chiang, E., Kite, E., Kalas, P., Graham, J. \& Clampin, M. 2009, ApJ 693, 734]
\item[Currie, T., Plavchan, P. \& Kenyon, S. 2008, ApJ 688, 597]
\item[Davis, D. S., Richer, H., Rich, R. M., Reitzel, D. \& Kalirai, J. 2009, ApJ 705, 398]
\item[Debes, J. \& Sigurdsson, S. 2002, ApJ 572, 556]
\item[Desharnais, S., Wesemael, F., Chayer, P., Kruk, J. \& Saffer, R. 2008, ApJ 672, 540]
\item[Dufour, P. et al. 2007, ApJ 663, 1291]
\item[Dufour, P. et al. 2010, ApJ in press (arXiv1006.3710)]
\item[Duncan, M. \& Lissauer, J. 1998, Icarus 134, 303]
\item[Dupuis, J,, Fontaine, G., Pelletier, C., \& Wesemael, F. 1992, ApJS 82, 505]
\item[------------ 1993a, ApJS 84, 73]
\item[Dupuis, J., Fontaine, G. \& Wesemael, F. 1993b, ApJS 87, 345]
\item[Eisenstein, D. et al. 2006, ApJS 167, 40]
\item[Farihi, J., et al. 2010b, MNRAS in press (arXiv:1001.5025)]
\item[Farihi, J., Jura, M., Lee, J-E \& Zuckerman, B. 2010a, ApJ 714, 1386]
\item[Farihi, J., Jura, M. \& Zuckerman, B. 2009, ApJ 694, 805]
\item[Gansicke, B., Marsh, T., Southworth, J. \& Rebassa-Mansergas, A. 2006, Science 314, 1908]
\item[Gorlova, N. et al. 2007, ApJ 670, 516]
\item[Jura. M. 2003, ApJ, 548, L91]
\item[Jura, M. 2006, ApJ, 653, 613]
\item[Jura, M. 2008, AJ, 135, 1785]
\item[Jura, M., Farihi, J. \& Zuckerman, B. 2007a, ApJ, 663, 1285]
\item[Jura, M., Farihi, J., Zuckerman, B. \& Becklin E. E. 2007b, AJ 133, 1927]
\item[Jura, M., Farihi, J. \& Zuckerman, B. 2009a, AJ 137, 3191]
\item[Jura, M., Muno, M. P., Farihi, J. \& Zuckerman, B. 2009b, ApJ 699, 1473]
\item[Jura, M. \& Xu, S. 2010, AJ in press (arXiv:1001.2595)]
\item[Kenyon, S., Shipman, H. Sion, E. \& Aannestad, P. 1988, ApJ 328, L65]
\item[Kilic, M., Farihi, J., Nitta, A. \& Leggett, S. K. 2008, AJ 136, 111]
\item[Kilic, M. \& Redfield, S. 2007, ApJ 660, 641]
\item[Kilic, M. von Hippel, T. Leggett, S. K. \& Winget, D.E. 2006, ApJ 646, 474]
\item[Klein, B., Jura, M., Koester, D., Zuckerman, B. \& Melis, C. 2010, ApJ 709, 950.]
\item[Koester, D.  2009, A\&A 498, 517]
\item[Koester, D. 2010, Mem. S.A.I. in press (eprint arXiv:0812.0482)]
\item[Koester, D. \& Wilken, D. 2006, A\&A 453, 1051]
\item[Koester, D. et al. 2005, A\&A 432, 1025]
\item[Limoges, M. \& Bergeron, P. 2010, ApJ 714, 1037]
\item[Lodders K. 2003, ApJ 591, 1220]
\item[Macintosh, B. et al. 2008, Proc. SPIE, 7015, 31]
\item[Marois, C. et al 2008, Science 322, 1348]
\item[McCook, G. \& Sion, E. 1999, ApJS 121, 1]
\item[Melis, C. et al. 2010, submitted to ApJ.]
\item[Osterbrock, D. 1974, Astrophysics of Gaseous Nebulae (W.H. Freeman, San Francisco)]
\item[Reach, W. et al 2005, ApJ 635, L161]
\item[Rebull, L. et al. 2008, ApJ 681, 1484]
\item[Rhee, J., Song, I., Zuckerman, B. \& McElwain, M. 2007, ApJ 660, 1556]
\item[Sfeir, D., Lallement, R., Crifo, F. \& Welsh, B. 1999, A\&A 346, 785]
\item[Sion, E., Holberg, J., Oswalt, T., McCook, G. \& Wasatonic, R. 2009, AJ 138, 1681]
\item[Su, K. et al. 2006, ApJ 653, 675]
\item[Su, K. et al. 2009, ApJ, 705, 314]
\item[Tremblay, P.-E. \& Bergeron, P. 2008, ApJ 672, 1144]
\item[Trilling, D. et al. 2008, ApJ 674, 1086]
\item[Trujillo, C., Jewitt, D. \& Luu, J. 2001, AJ 122, 457]
\item[Vennes,S., Kawka, A. \& Nemeth, P. 2010, MNRAS in press (astroph 1002.2069)]
\item[Vogt, S. et al. 1994, SPIE 2198, 362]
\item[von Hippel, T, Kuchner, M., Kilic, M., Mullaly, F. \& Reach, W. 2007, ApJ 662, 544]
\item[Voss, B., Koester, D., Napiwotski, R., Christlieb, N. \& Reimers, D. 2007, A\&A 470, 
1079]
\item[Wolff, B., Koester, D. \& Liebert, J. 2002, A\&A 385, 995]
\item[Zuckerman, B., Koester, D., Reid, I. N. \& Hunsch, M. 2003, ApJ 596, 477 (ZKRH03)]
\item[Zuckerman, B., Koester, D., Melis, C., Hansen, B. \& Jura, M. 2007, ApJ 671, 872]
 
\end{harvard}

\clearpage
\begin{figure}
\includegraphics[width=140mm]{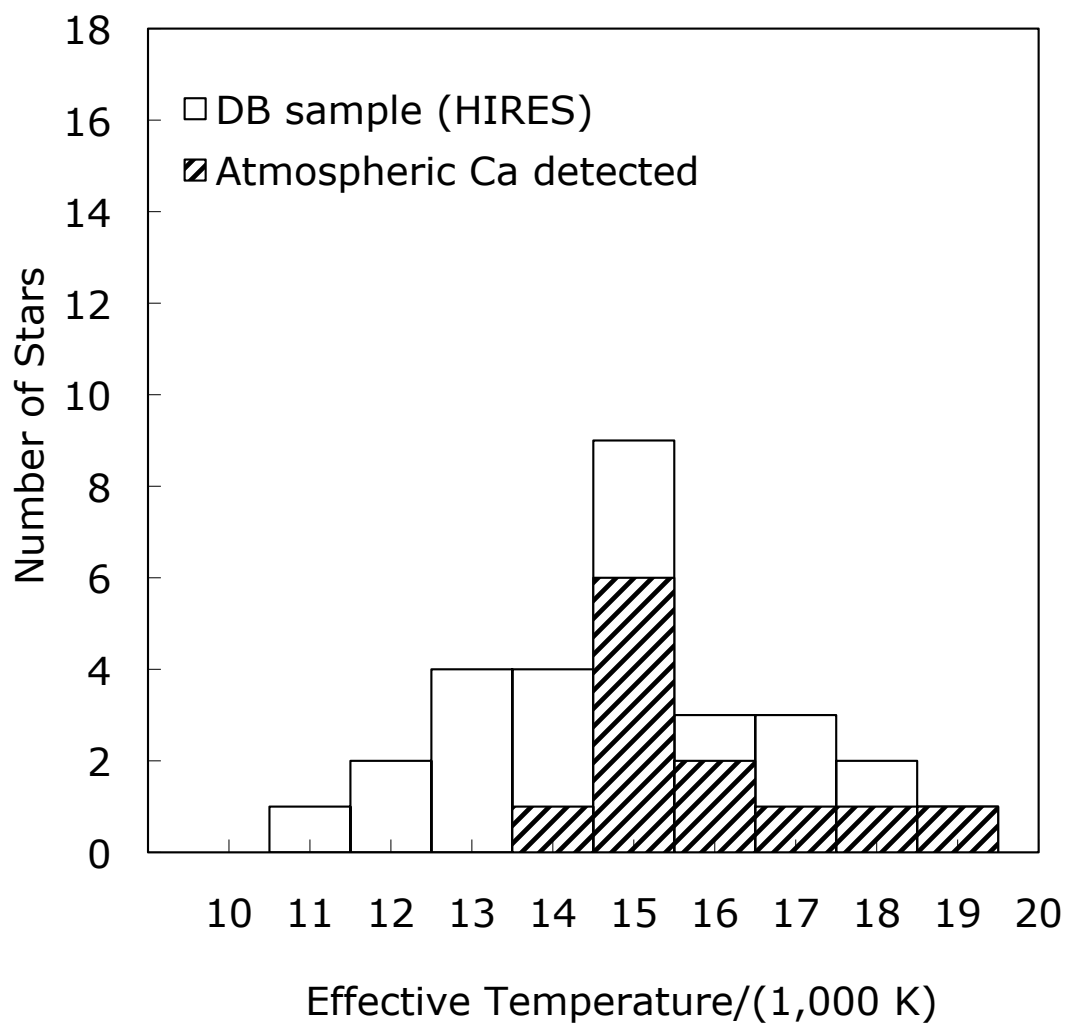}
\caption{\label{figure1} Histogram displaying the number of white dwarfs from Table 1 with photospheric calcium (lined rectangles) and without calcium (open rectangles) vs. stellar effective temperature.  There are 12 stars with Ca and 18 without.}
\end{figure}

\clearpage
\begin{figure}
\includegraphics[width=140mm]{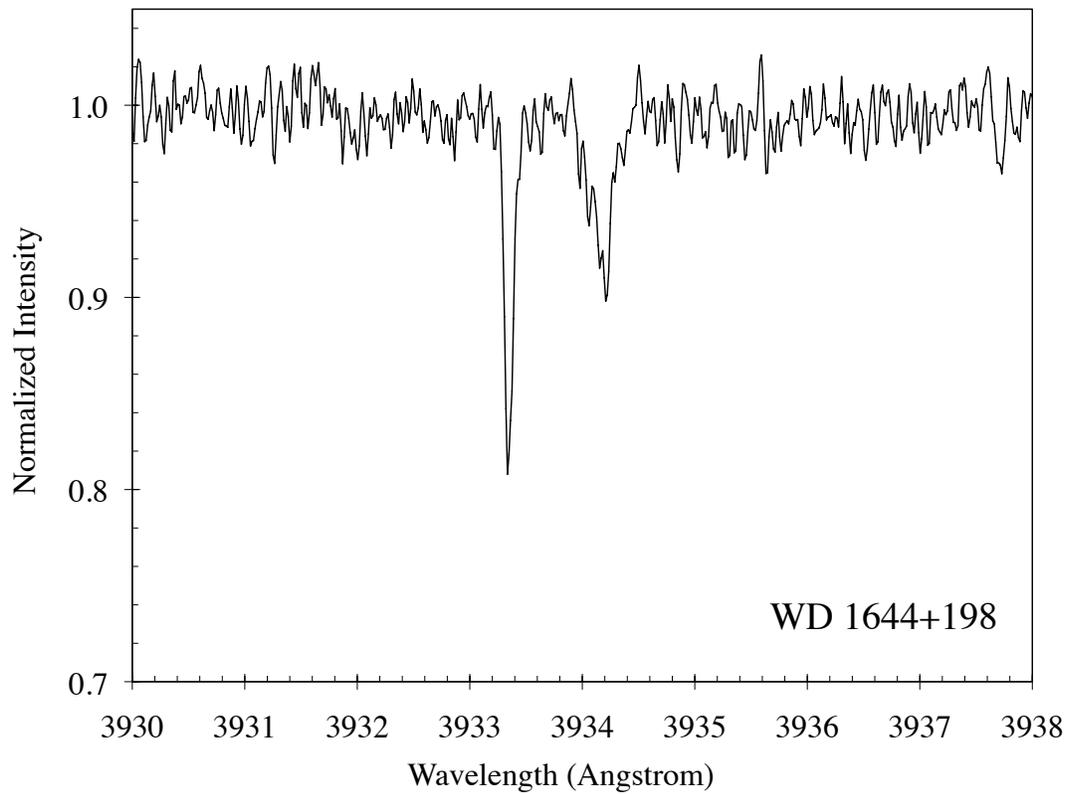}
\caption{\label{figure 2} HIRES spectrum of WD1644+198 in the vicinity of the Ca II K-line.  The redshifted K-line absorption feature is from the photosphere and the blueshifted one is interstellar in origin.  The abscissa is wavelength in air in the heliocentric rest frame.}
\end{figure}

\clearpage
\begin{figure}
\includegraphics[width=140mm]{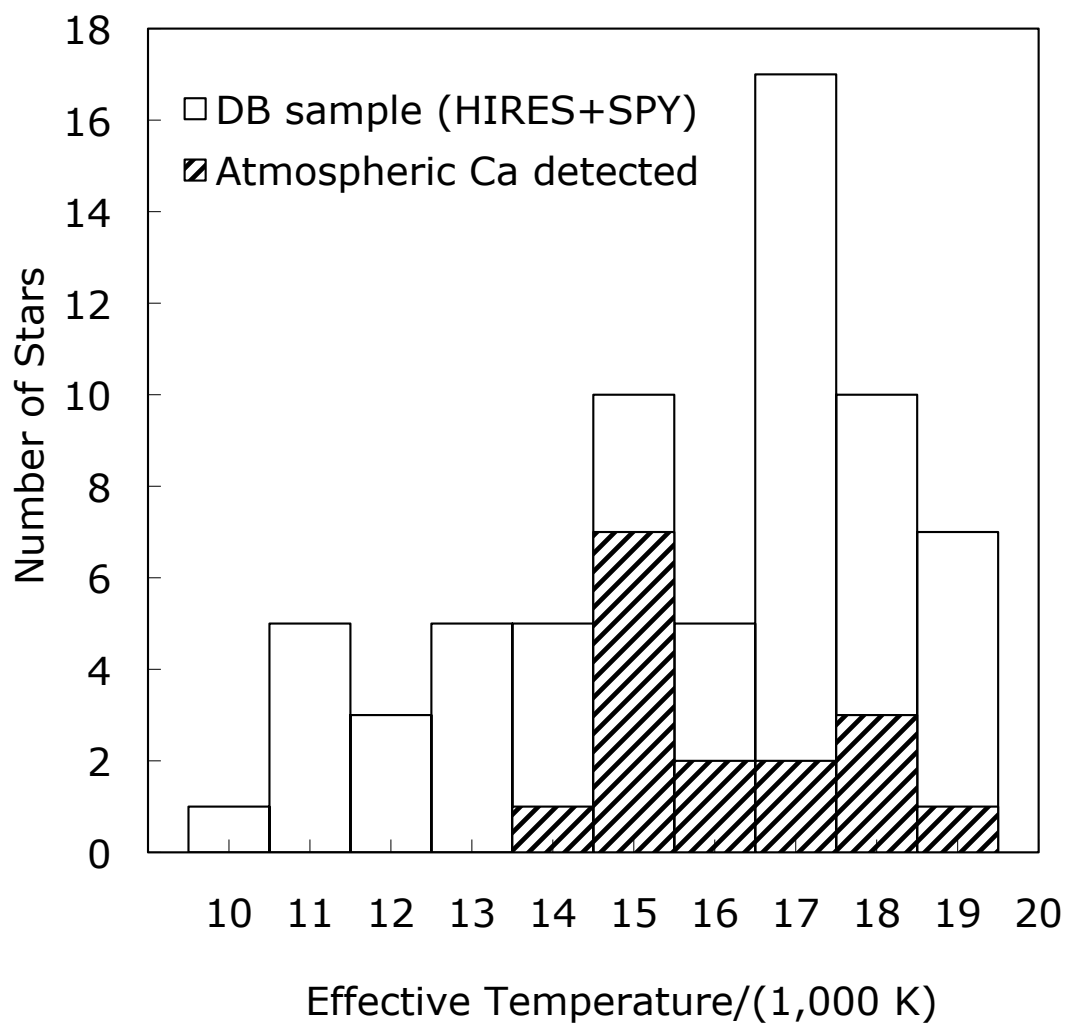}
\caption{\label{figure 3} Histogram displaying the number of white dwarfs from Table 1 and from Voss et al (2007) with photospheric calcium (lined rectangles) and without calcium (open rectangles) vs. stellar effective temperature.  WD2229+139 is included as a DBAZ (as indicated in Koester et al 2005, but not in the comments column of Table 2 of Voss et al 2007).  WD1134+073 is indicated as a DBAZ in Voss et al (2007) but upon comparison of the Ca II K-line radial velocity with the velocities of some strong He lines, we find the weak Ca line to be strongly blueshifted with respect to the He lines.  Therefore, the K-line is probably interstellar and we classify 1134+073 as a DB star.}
\end{figure}

\clearpage
\begin{figure}
\includegraphics[width=140mm]{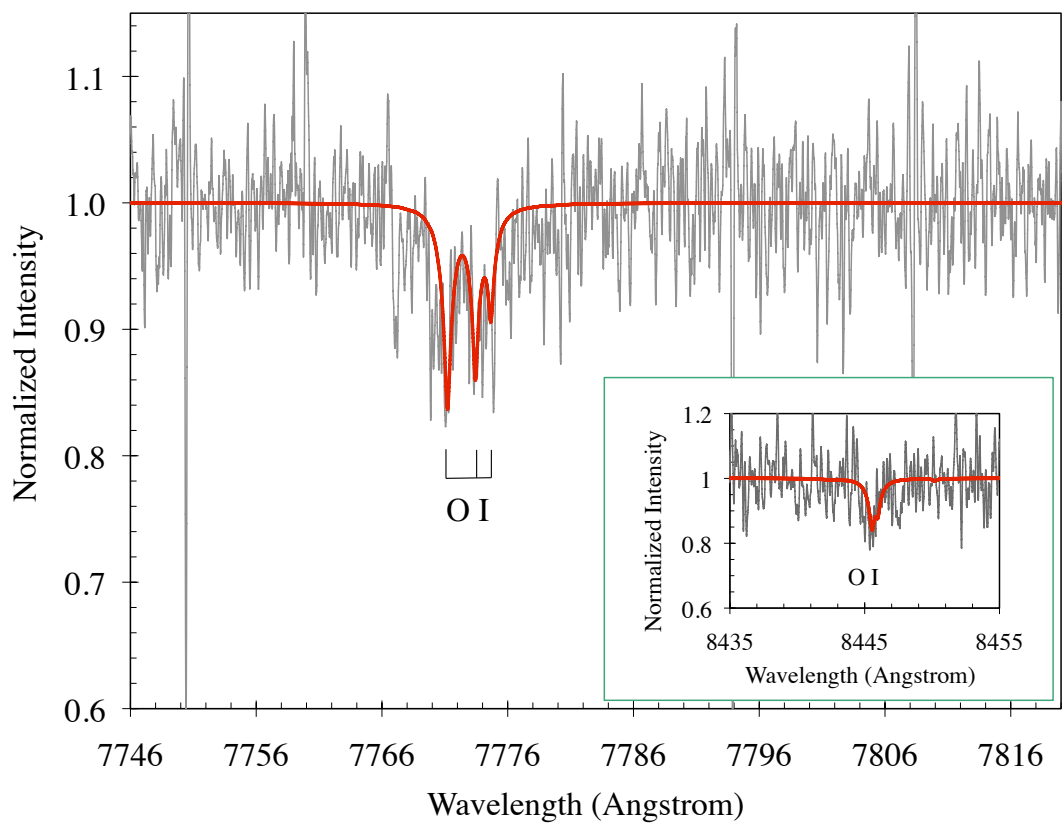}
\caption{\label{figure 4} Spectrum of DBZ white dwarf G-241-6.  These portions of a 3800 sec HIRES exposure display weak OI lines. The data are smoothed with a 7 point boxcar average.  Wavelengths are in air and a heliocentric frame of rest. Several sky lines (e.g. near 7750 and 7794 \AA) were not easily removed in sky subtraction and remain as narrow, spiked absorption/emission features.  The model spectrum (thick red line) with Teff = 15,300 K and log g = 8.0 represents an oxygen abundance of log[n(O)/n(He)] = -5.6, and is shown blue shifted by -28.0 km/s, the mean velocity from the full set of measured absorption lines.}
\end{figure}

\clearpage
\begin{table}
\caption{Atmospheric properties of DB white dwarfs}
\begin{tabular}{@{}lccccccc}
\br
WD& Name& Type& T$_{eff}$& EW& [Ca/He]& log(dM/dt)& t$_{cool}$\\
& & & (K)& (m\AA)& & (g/s)& (10$^8$ yr)\\
\mr
0000-170& G266-32& DB& 12,900& $<$12& $<$-11.3& $<$5.86& \\
0125-236& G274-39& DBAZ& 16,400& 29$\pm$5& -9.4& 7.50& 1.6\\
0129+246& PG& DB& 14,800& $<$35& $<$-10.0& $<$7.08& \\
0517+771& GD435& DBA& 13,250& $<$11& $<$-11.2& $<$5.92& \\
0845-188& L0748-70& DB& 17,450& $<$15& $<$-9.2& $<$7.53& \\
1046-017& GD124& DB& 14,300& $<$9& $<$-10.9& $<$6.20& \\
1056+345& G119-47& DB& 12,000& $<$8& $<$-11.9& $<$5.29& \\
1107+265& GD128& DBZ& 14,850& 45$\pm$4& -9.9& 7.17& 2.2\\
1129+373& PG& DB& 12,800& $<$24& $<$-11.1& $<$6.07& \\
1352+004& PG& DBAZ& 14,200& 112$\pm$11& -9.7& 7.37& 2.6 \\
1403-010& G64-43& DBA& 15,600& $<$10& $<$-10.6& $<$6.44& \\ 
1411+218& PG& DB& 15,000& $<$5& $<$-10.8& $<$6.26& \\
1459+821& G256-018& DB& 15,000& $<$7& $<$-10.7& $<$6.36& \\
1545+244& Ton249& DB& 12,700& $<$23& $<$-11.1& $<$6.07& \\
1610+239& PG& DB& 14,000& $<$9& $<$-11.0& $<$6.11& \\
1644+198& PG& DBZ& 15,000& 17$\pm$1& -10.3& 6.76& 2.2 \\
1709+230& GD205& DBAZ& 19,250& 45$\pm$4& -7.9& 8.73& 0.88 \\
1940+374& L1573-031& DB& 16,900& $<$5& $<$-10.1& $<$6.73& \\
2058+342& GD392& DB& 11,900& $<$13& $<$-11.8& $<$ 5.37& \\
2129+000& G26-10& DB& 14,000& $<$7& $<$-11.1& $<$6.01& \\
2130-047& GD233& DBA& 18,200& $<$8& $<$-9.2& $<$7.40& \\
2144-079& GL837.1& DBZ& 16,000& 132$\pm$13& -8.6& 8.08& 1.7 \\
2147+280& G188-27& DB& 11,000& $<$8& $<$-12.4& $<$4.84& \\
2222+683& G241-6& DBZ& 15,300& 960$\pm$96& -7.25& 9.30& 1.9 \\
2234+064& PG& DB& 21,500& $<$15& $<$-8.0& $<$7.20& \\
& & & & & & & \\
0002+729& GD408& DBZ& 14,200& & -9.3& 7.83& 2.7 \\
0300-013& GD40& DBAZ& 15,300& 2500$\pm$200& -6.9& 9.44& 1.9 \\
1011+570& GD303& DBZ& 18,000& & -7.75& 8.84& 0.76 \\
1425+540& G200-039& DBAZ& 14,750& 150$\pm$50& -9.3& 7.73& 2.2 \\
1822+410& GD378& DBAZ& 16,800& & -8.0& 8.58& 1.5 \\
\br
\end{tabular}
\end{table}
\noindent Note $-$ EW is for the Ca II K-line.  [Ca/H] is the logarithm of the calcium to helium ratio by number.  Calcium data for the 25 stars above the blank line are from HIRES observations, for the five stars below the blank line from the literature (Klein et al. 2010; Wolff et al. 2002; Descharnais et al. 2008; Kenyon et al. 1988).   WD1403-010, 1644+198, 1940+374, and 2234+064 all display Ca K-line absorption from the interstellar medium (see Section 3).  WD1352+004, 1709+230, 2144-079, and 0300-013 show no evidence for excess near-IR emission out to 2.5 $\mu$m (Kilic et al. 2008).  dM/dt are current mass accretion rates of all high-Z elements based on assumption of steady-state accretion (see Section 4.1, Appendix B, and references listed in these Notes above).   Cooling times are from Bergeron et al. (1995).  Masses of all listed stars are assumed or known to be close to 0.58 M$_{\odot}$ except for GD303 for which we use a mass of 0.5 M$_{\odot}$ to find its cooling time. Limoges \& Bergeron (2010) derive an average DB mass of 0.76
M$_{\odot}$. If the mass of a typical (15,000 K) DBZ star in our survey
actually is 0.76 M$_{\odot}$ rather than 0.58 M$_{\odot}$, then [Ca/He] would be
about 25\% larger than listed in Table 1.

\clearpage
\begin{table}
\begin{indented}
\caption{  DB white dwarfs: mass accretion rate vs. cooling age}
\item[]\begin{tabular}{@{}lcc}
\br
DB sample from Table 1& median t$_{cool}$& median log (dM/dt)\\
& (10$^8$ yr)& (g/s)\\
\mr
7 hottest DBZ& 1.6& 8.73\\
12 hottest (DB+DBZ)& 1.6& 7.79\\
5 coolest DBZ& 2.4& 7.37\\
13$^{th}$-23$^{rd}$ hottest (DB+DBZ)& 2.4& 6.7\\
7 coolest (all DB)& 3.55& $<$5.86\\
\br
\end{tabular}
\end{indented}
\end{table}
\indent Note $-$ The white dwarfs in the first two entries have effective temperatures equal to or greater than that of GD40 (15,300 K).  The next two entries lie in the temperature range 14,000 to 15,000 K.  The last entry contains DB stars with T$_{eff}$ $<$13,500 K (the three lowest temperature bins in Fig. 1). 

\clearpage
\begin{table}
\begin{indented}
\caption{ Distribution of spectral types of white dwarfs within 20 pc}
\item[]\begin{tabular}{@{}lcc}
\br
Spectral Type& Number of stars& \% of total\\
\mr
DP, DH& 17& 13\\
DA& 60& 47\\
DAZ& 8& 6\\
DZ& 9& 7\\
DQ& 12& 9\\
DC& 22& 17\\
DBQA& 1& 1\\
\br
\end{tabular}
\end{indented}
\end{table}
\indent Note $-$ Many stars in this sample of 129 have not been observed with a sensitive spectroscopic system (see discussion in Section 4.3.2).   This table is a revision of Table 2 in Sion et al. (2009).

\begin{deluxetable}{lccccccccccc}
\rotate
\centering
\tabletypesize{\scriptsize}
\tablecolumns{12}
\tablewidth{0pt}
\tablecaption{Kinematics of DB White Dwarfs \label{tabkin}}
\tablehead{
 \colhead{WD Name} & 
 \colhead{RA} & 
 \colhead{DEC} & 
 \colhead{RV$_{\rm obs}$} & 
 \colhead{v$_{\rm corr}$} & 
 \colhead{Distance} & 
 \colhead{pmRA} &
 \colhead{pmDE} & 
 \colhead{U} & 
 \colhead{V} & 
 \colhead{W} &
 \colhead{Total} \\
 \colhead{} & 
 \multicolumn{2}{c}{(J2000)}& 
 \colhead{(km s$^{-1}$)}  & 
 \colhead{(km s$^{-1}$)}  & 
 \colhead{(pc)}     & 
 \colhead{(mas yr$^{-1}$)} & 
 \colhead{(mas yr$^{-1}$)} & 
 \colhead{(km s$^{-1}$)} & 
 \colhead{(km s$^{-1}$)} & 
 \colhead{(km s$^{-1}$)} &
 \colhead{}
}
\startdata
G 266$-$32    & 00 03 31.58 & $-$16 43 58.5 & +28$\pm$10     & $-$2 & 41  & +245 & $-$7 & $-$42 & $-$23 & $-$7 & 48 \\
G 274$-$39    & 01 27 44.55 & $-$23 24 48.6 & +59$\pm$5       & +29 & 75       &  +302 & +44 & $-$99 & $-$53 & $-$12 & 113  \\
PG 0129+246 & 01 32 24.06 & +24 56 13.0    & +17$\pm$10    & $-$13 & 102 & $-$38 & $-$38 & +27 & $-$7 & $-$10 & 29 \\
GD 435            & 05 24 59.03 & +77 14 14.9    & $-$13$\pm$10 & $-$43 & 88   & +90 & $-$63 & +6 & $-$62 & 4 & 63 \\
L 0748$-$70   & 08 47 29.45 & $-$18 59 49.8 & +99$\pm$10     & +69 & 95      & $-$190 & +38 & $-$94 & $-$45 & $-$39 & 111 \\
GD 124            & 10 48 32.66 & $-$02 01 10.8 & +11$\pm$10     & $-$19 & 75   & $-$6 & $-$128 & +24 & $-$22 & $-$37  & 49 \\
G 119$-$47    & 10 59 25.54 & +34 14 52.7     & +29$\pm$10     & $-$1 & 54   & $-$187 & $-$219 & $-$24 & $-$68 & $-$17 & 74 \\
GD 128            & 11 09 59.83 & +26 18 47.7     & $-$12$\pm$5   & $-$42 & 78   & +79 & $-$133 & +58 & $-$28 & $-$30 & 71 \\
PG 1129+373 & 11 31 43.37 & +37 01 28.6    & $-$6$\pm$10   & $-$36 & 93   & +92 & $-$24 & +51 & +5 & $-$20 & 55 \\
PG 1352+004 & 13 55 32.40 & +00 11 23.5    & +26$\pm$5       & $-$4 & 75   & $-$38 & $-$38 & $-$5 & $-$18 & $-$6 & 20 \\
G 64$-$43       & 14 06 20.01 & $-$01 19 32.5  & +2$\pm$10       & $-$28 & 81   & $-$243 & $-$91 &  $-$65& $-$80 & $-$13 & 104 \\
PG 1411+218 & 14 13 29.85 & +21 37 39.8      & $-$18$\pm$10 & $-$48 & 43   & $-$50 &+108  & $-$36 & +4 & $-$40 & 54 \\
G 256$-$018  & 14 56 31.89 & +81 56 52.4      & +64$\pm$10     & +34 & 53      & $-$327 & +167 & $-$91 & $-$23 & +28  & 98 \\
Ton 249           & 15 47 54.49 & +24 20 39.8      & +2$\pm$10       & $-$28 & 65   & $-$50 & +65 & $-$36 & $-$9 & $-$9 & 38 \\
PG 1610+239 & 16 13 01.54 & +23 48 30.8      & +18$\pm$10    & $-$12 & 65   & +80 & $-$18 & +6 & +7 & $-$26 & 25 \\
PG 1644+198 & 16 46 19.76 & +19 46 03.7      & +40$\pm$5       & +10 & 59     & +54 & $-$94 & +29 & $-$2 & $-$13 & 30  \\
GD 205            & 17 11 55.66 & +23 01 01.9      & +47$\pm$5       & +17 & 69         & +14 & $-$164 & +53 & $-$17 & $-$10 & 56 \\
L 1573$-$031 & 19 42 13.00 & +37 31 56.4     & +45$\pm$10    & +15 & 50         & $-$16 & +225 & $-$39 & +25 & +31 & 56 \\
GD 392            & 21 00 21.51 & +34 26 20.9     & +22$\pm$10     & $-$8 & 57   & +122 & +120 & $-$47 & +1 & $-$3 & 47 \\
G 26$-$10       & 21 32 16.24 & +00 15 14.4     & $-$84$\pm$10 & $-$114 & 40 & +416 & +34 & $-$115 & $-$76 & +15 & 77  \\
GD 233            & 21 33 34.83 & $-$04 32 24.2  & +10$\pm$10    & $-$20 & 55    & +241 & +13 & $-$58 & $-$13 & $-$29  & 66 \\
GL 837.1         & 21 47 37.25 & $-$07 44 12.2  & $-$2$\pm$5      & $-$32 & 52    & +250 & $-$132 & $-$49 & $-$51 & $-$30 & 77  \\
G 188$-$27    & 21 49 54.53 & +28 16 59.8      & $-$15$\pm$10 & $-$45 & 35    & +250 & $-$84 & $-$30 & $-$51 & $-$21 & 63 \\
G 241$-$6       & 22 23 33.08 & +68 37 24.2     & $-$28$\pm$5   & $-$58 & 72    & +143 & +242 & $-$59 & $-$90 & +33 & 112  \\
PG 2234+064 & 22 36 41.99 & +06 40 17.2     & +21$\pm$10    & $-$9 & 125  & +22 & 0  & $-$13 & $-$10 & 0  & 16 \\
\enddata
\tablecomments{U, V,W are in km s$^{-1}$ with respect to the Sun and are positive in the directions of the Galactic center, Galactic rotation, and the north Galactic pole.  v$_{corr}$ assumes a gravitational redshift velocity  = +30 km s$^{-1}$.  Uncertainties in distance between Earth and a white dwarf are assumed to be $\sim$10\%, and in each component of proper motion to be $\sim$5 mas yr$^{-1}$. The uncertainties in each component of UVW are typically about 5-10 km/s.
Total = $\sqrt{U^2 + V^2 + W^2}$.} 
\end{deluxetable}

\clearpage
\begin{table}
\begin{indented}
\caption{ G241-6: atmospheric abundances and mass accretion rates}
\item[]\begin{tabular}{@{}lcccc}
\br
Z& \#& log[Z/He]& $\pm$err& log(dM/dt)\\
& & & & (g/s)\\
\mr
O& 4& -5.6& 0.10& 8.96\\
Mg& 7& -6.29& 0.05& 8.55\\
Si& 7& -6.78& 0.06& 8.13\\
Ca& 9& -7.25& 0.07& 8.00\\
Ti& 12& -8.97& 0.04& \\
Cr& 4& -8.46& 0.05& \\
Mn& 1& -8.8& 0.10& \\
Fe& 9& -6.76& 0.06& 8.70\\
H& & $<$-6.1& & \\
Al& & $<$-7.0& & \\
Ni& & $<$-7.5& & \\
\br
\end{tabular}
\end{indented}
\end{table}
\indent Note $-$ The listed elemental abundances are the total in all stages of ionization; for Si, Ca, Ti, Cr, Mn and Fe they are based on transitions of the singly ionized species.  The oxygen abundance is from transitions of OI (see Fig. 4).  The Mg abundance is from four transitions in Mg I and three in Mg II; these agree at the listed value of [Mg/H] for an atmosphere model with T$_{eff}$ = 15,300 K and log g = 8.  Mg, Si, Ca, Ti and Fe errors are mean deviations of the set of lines.  O, Cr, and Mn errors are dominated by measurement uncertainties.  Additional systemic uncertainties arise from uncertainty in the atmospheric model temperature and gravity and probably amount to an additional 0.1 dex error.   dM/dt are current mass accretion rates based on the assumption of steady-state accretion and have been calculated for only the 5 most abundant of the detected elements.  While plausible uncertainties in log g can substantially impact the mass
of the convection zone (e.g., Table 2 in Dufour et al. 2010), such
uncertainties have much less of an effect on dM/dt and it is dM/dt that is
a major focus of the present paper.

\end{document}